# Grid Inadequacy Assessment against Power Injection Diversity from Intermittent Generation, Dynamic Loads, and Energy Storage

Adonis E. Tio, David J. Hill, *Life Fellow, IEEE* and Jin Ma

*Abstract* — **The integration of more intermittent generation, energy storage, and dynamic loads on top of a competitive market environment requires future grids to handle increasing diversity of power injection states. Grid planners need new tools and metrics that can assess how vulnerable grids are against this future. To this end, we propose grid inadequacy metrics that expose a grid's inability to accommodate power injection diversity from such sources. We define the metrics based on a previously unexplored characterization of grid inadequacy, that is, the size of the DC power flow infeasible set relative to the size of the power injection set is indicative of inherent grid inadequacy to accommodate power injection diversity without intervention. We circumvent the difficulty of characterizing the high-dimensional sets involved using three approaches: one sampling-based approach and two approaches that project the sets in lower dimensions. Illustrative examples show how the metrics can reveal useful insights about a grid. As with other metrics, the proposed metrics are only valid relative to the assumptions used and cannot capture all intricacies of assessing grid inadequacy. Nevertheless, the metrics provide a new way of quantifying grid inadequacy that is potentially useful in future research and practice. We present possible use-cases where the proposed metrics can be used.**

## I. INTRODUCTION

The integration of bulk and distributed intermittent renewable generation, dynamic loads, and energy storage systems on top of a competitive market environment is expected to introduce more diversity in power injection states in future grids. Coupled with long-term uncertainty in user demand patterns, demographic trends, climate change, economic activity, and policies around new technologies, it will increasingly become more difficult to design adequate, secure, reliable, and cost-effective grids. There is a need to develop new measures of grid adequacy, inadequacy, and risks that captures the interrelated uncertainties in both operational and strategic timescales.

This work looks at a new approach to defining grid inadequacy in future grids with increased diversity in power injection states. We characterize grid inadequacy as a grid's inherent inability to serve all possible power injection states without any intervention. The proposed inadequacy metrics are based on the following idea that we believe is not yet considered in existing literature. That is, given all possible power injection scenarios, the proportion of infeasible scenarios that lead to network congestion provide an indicative measure of inherent grid inadequacy against power injection diversity. This work provides an initial working framework that explores this characterization of grid inadequacy that researchers and practitioners may find useful in the future.

In the proposed framework, we represent the uncertainty in bus power injections from renewables, dynamic loads, and energy storage together with injections from more conventional generation and loads using interval sets that span the whole power injection space. We then use the constrained DC power flow model to identify infeasible subspaces. Since it is generally difficult to make sense of the high-dimensional sets involved, we use three approaches to develop numeric quantities or metrics that describe the infeasible set. While single numeric values cannot fully capture the complex relationships between the sets involved, let alone capture the intricacies of defining grid inadequacy, the metrics can still provide valuable insights on network vulnerabilities and bottlenecks that can motivate detailed analyses later on.

We organize the paper as follows. Section II reviews the available literature on power injection diversity and grid adequacy assessment, identifies shortcomings, and motivates the need for new dedicated measures of grid inadequacy in the future grid setting. Section III presents a framework for measuring grid inadequacy using the infeasible set of the constrained DC power flow model. Section IV defines tractable metrics that embody the framework presented in Section III. Section V show examples for the 6- and 118-bus test systems. Section VI comments on the framework assumptions, identifies areas for future work, and presents a potential use-case. Section VII concludes.

A. E. Tio acknowledges the DOST-ERDT Faculty Development Fund of the Republic of the Philippines for sponsoring his degree leading to this research.

A. E. Tio is a PhD student at the School of Electrical and Information Engineering, University of Sydney, Sydney, NSW 2006, Australia and a faculty (on-leave) of the Electrical and Electronics Engineering Institute, University of the Philippines, Diliman 1101, Quezon City, Philippines (e-mail: adonis.tio@eee.upd.edu.ph).

D. J. Hill is with the Department of Electrical and Electronics Engineering, The University of Hong Kong, Hong Kong, and the School of Electrical and Information Engineering, University of Sydney, Sydney, NSW 2006, Australia (e-mail: dhill@eee.hku.hk).

J. Ma is with the School of Electrical and Information Engineering, University of Sydney, Sydney, NSW 2006, Australia (e-mail: j.ma@sydney.edu.au).

## II. POWER INJECTION DIVERSITY AND GRID INADEQUACY

### A. Grid adequacy assessment

Ref. [1] defines power system adequacy as the existence of enough facilities to meet customer needs and operational constraints at static conditions under a set of probabilistic system states. Assessing the adequacy of the transmission network is generally complicated because of complexities in both (a) network and operations modeling and (2) system state selection, see Fig. 1. Network and operations modeling involves the choice of a power flow model, whether contingencies are considered, and whether power flow control such as power injection, topology, impedance, or voltage angle control are modeled. System state selection depends on the application and can span both probable and possible system states hours or days ahead in an operational planning setting and years or decades ahead in a strategic planning setting. System states are much more predictable in the operational setting with future system states more likely to resemble system states from the immediate past. On the other hand, system states in the distant future are much harder to predict because of the dependency on extraneous factors such as climate change, technological development, and legal and market restructuring. Because of this complexities, measures of grid adequacy are highly dependent on the assumptions used and applicable only within the intended use-case.

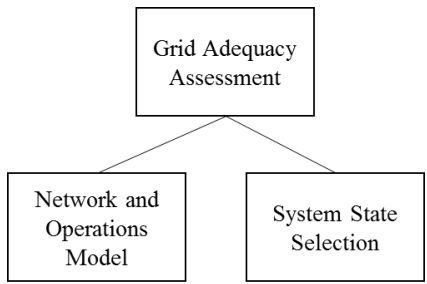

Fig. 1 Basic components of grid adequacy assessment.

With the integration of more intermittent renewable generation, dynamic loads, and energy storage on top of existing deregulated electricity grids, the number and diversity of system states are expected to increase. Buses can both become a source and a sink as energy storage becomes embedded in generator buses with either dispatchable or intermittent generation as well as in load buses with distributed generation, traditional loads, and dynamic loads. Combined with new market, data, and communications infrastructure, market players will have more and more flexibility in controlling when and how much energy they will draw or inject from the grid depending on prevailing and forecasted market conditions. These trends will require a rethink of what grid adequacy means. Ideally, future grids must be able to facilitate all possible energy exchanges between future market players without discrimination, while keeping the service secure, reliable, and low-cost. New tools and metrics must be explored to study the adequacy of existing grids and proposed reinforcements against such a future.

This work looks at the antithesis of grid adequacy which is grid inadequacy. As with characterizing adequacy, characterizing grid inadequacy is complicated and highly depends on the modeling assumptions used and system states modeled. It is equally important to quantify grid inadequacy (what a grid cannot do) but most work tends to focus on characterizing grid adequacy (what a grid can do). Since doing grid adequacy assessment does not necessarily provide grid inadequacy information, there is a need to study and develop dedicated assessment frameworks and measures of grid inadequacy.

### B. Existing metrics

We first review the generic adequacy metrics presented in [1] that can be used with any combination of modeling assumptions used and system states chosen. We then limit our literature review to more application-specific measures of grid adequacy under non-contingency situations against a defined set of power injection states or scenarios. The papers use the DC power flow model and most consider power flow control interventions in the form of generator dispatch flexibility or load shedding but not topology, impedance, or voltage angle control. The range of power injection states reflect the intended applications and spans hours, days, years, or decades ahead into the future.

Ref. [1] presents seven basic grid adequacy indices and five derived indices. The indices revolve around characterizing load curtailments given a discrete pool of system states. Some of the more notable ones include the *Probability of Load Curtailments (PLC)*, *Expected Number of Load Curtailments (ENLC)*, and *Expected Energy Not Supplied (EENS)*. *PLC* is the sum of probabilities of all states that lead to curtailment while *ENLC* is the total number of load curtailment states. *EENS* gives the expected energy curtailment within a given period. It is more commonly encountered in generation adequacy assessment but can also be used to incorporate the effects of transmission adequacy. These metrics are defined broadly enough to accommodate any level of network and operations modeling fidelity over a range of pre-selected system states. The basic framework is to enumerate discrete system states and evaluate which states lead to curtailment. After evaluating the enumerated states, adequacy metrics can be used to describe the properties of the curtailment states. The obvious drawback of this framework is that the metrics are only as good as the system states identified. As such, proper context and caveats must be established when using these metrics in decision-making.

Other works define measures of grid adequacy more specifically and tailored to an application. In works related to operations planning, measures of grid adequacy are described in terms of security, flexibility, loadability, or robustness. Despite the different terminologies used, these measures pertain to a grid's ability to accommodate power injection states under static conditions, in contrast to other metrics evaluated under dynamic conditions. Some work characterize adequacy in terms of sets while others define metrics. In [2]–[4], security regions are used to define allowable variations in power injections about an operating point. To compute the allowable adjustments, Ref. [2] uses a linear program based on the constrained DC power flow model to maximize the sum of allowable adjustment range of all generators. A performance index is then defined using the proportion of the allowable generation adjustment relative to the maximum generation adjustment for the whole system. Likewise, Ref. [5] uses the allow-able generator adjustments to define grid flexibility but solves for the ranges using a different approach: by taking the difference between a reference dispatch and a new dispatch that minimizes the line loading margins. A flexibility index is defined as the maximum generation variation in each generator bus or the whole system. Ref. [6] uses the range of allowable load variations to define grid robustness under load uncertainty. The authors solve a linear program repeatedly to approximate the extent of allowable load changes about a loading condition. A robustness index is then defined in terms of the probability of a grid working correctly even if it operates outside a specified set of operating conditions. And in [7], a set of constraints is used to define loadability sets. The paper notes that metrics derived from the regions bounded by the loadability sets, like inner-area approximations and centroid calculations, would provide useful indicators of grid flexibility and are important areas of future research.

In works related to infrastructure planning, Refs. [8] and [9] use an ad hoc sampling-based approach to compare the robustness of grid expansion options to both load and renewable generation uncertainty. $K$ scenarios are drawn from the distribution of uncertain load and generation variables and the number of scenarios that do not result in curtailment, $K_1$, are counted. The larger the ratio of $K_1$ and $K$, the more robust a grid is to power injection variability. This approach follows the framework behind *ENLC* tailored for a planning problem under load and renewable generation uncertainty. Ref. [10] uses the maximum load shedding under uncertainty as metric for comparing grid alternatives. Ref. [11] defines a framework to quantify the benefits of grid-side flexibility options provided by line switches and FACTS devices using sets defined from the constrained DC power flow model. The idea is to compare the size of feasible sets with and without flexibility options installed. Instead of defining discrete system states, spaces in the power injection space are used to represent system states in a continuum. And in a completely radical perspective, Ref. [12] uses a graph theoretic approach to grid adequacy assessment. The authors propose to use the volume of feasible bus injections to measure grid flexibility to absorb and deliver power. The authors claim that line additions that maximize the determinant of the DC power flow susceptance matrix will also maximize the volume of feasible bus injections under certain assumptions. However, it remains unclear whether improvements in underlying graph structures of power grids will indeed translate to actual improvements in grid adequacy.

### C. *Need for new metrics*

As we mentioned in Section II-A, adequacy metrics are only as good as the assumptions and methodologies used. As such, each has its own limitations and intended use-cases. In this section, we comment on the limitations of the reviewed works to motivate the need for new dedicated measures of grid inadequacy against increased power injection diversity in future grids.

The basic metrics presented in  [1] rely on the enumeration of discrete power injection states. In the future, scenario sampling and scenario reduction techniques can be used to derive a tractable but representative number of power injection states. However, planners must also explore grid vulnerabilities out-side of the pre-selected states and quantify risk measures on being in these states. The methods in [2]–[6] are defined for a specific operating point and are limited to modeling either generation variability only or load variability only. While useful in their respective intended applications, overall measures of grid performance that are independent of a specific operating point and can simultaneously capture load and generation variability will be useful.

To address these issues, the ideas in [8]-[12] provide inspiration, but these works have limitations as well. The sampling-based approach in [8] and [9] suffer the same blindness to unsampled scenarios as with the metrics in [1]. Moreover, without appropriate scenario selection and scenario reduction methods, sampling-based approaches can easily become computationally prohibitive if the number of uncertain variables increase, which is likely in the future as power injections become much more controllable and flexible. Using the maximum load shedding as indicator of grid adequacy is also useful when used in the proper context but it fails to capture the extent of network vulnerability. Moreover, the model assumes flexibility in generation dispatch and as such, masks a grid's inherent limitations. While [11] provides a useful framework for quantifying adequacy in terms of spaces instead of discrete scenarios, it is unclear how to operationalize this framework especially for a large number of uncertain variables. A visual comparison of feasible sets used as an illustration in [11] is possible in 2D space for two uncertain variables but is generally difficult for more variables. And while the exploration of the relationship between graph properties and adequacy of power grids is interesting, the work is still in its seminal stages. In [12], the determinant of the susceptance matrix is used as a proxy for the volume of feasible power injection and no proposal was given to actually quantify the latter parameter. This makes it difficult to know how much grid adequacy improved and how much risk of congestion remains.

 In addition to the limitations identified, all works reviewed focus on characterizing and measuring adequacy As such, explicit measures of grid inadequacy, are lacking. Since characterizing grid adequacy does not necessarily give information about grid inadequacy, it is difficult to estimate the extent of risk of congestion and assess the efficacy of risk mitigation solutions without dedicated grid inadequacy metrics. The lack of grid inadequacy metrics also hinders the development of complementary optimization models that provide solutions that minimize this problem.

To address the gaps identified, this work offers the following novel contributions:

a.  We present a framework that is specifically designed to characterize and measure inherent grid inadequacy to accommodate power injection diversity in future grids.

b.  The framework uses a previously unexplored way of characterizing grid inadequacy that is potentially useful in future research and practice. That is, given all system states within the whole power injection space, the relative size of the infeasible subspace is indicative of inherent grid inadequacy to accommodate power injection diversity without intervention.

c.  Since the sets involved are generally high-dimensional and difficult to analyze, we present three approaches to

circumvent this difficulty. One is a sampling-based approach while the other two are novel dimension-reduction approaches.

d. We present three inadequacy metrics that result from the three approaches used in characterizing the multidimensional sets.

The advantages of the proposed framework and metrics complement the limitations of the different approaches reviewed as follows. The metrics can capture the spatiotemporal diversity of both positive and negative power injections simultaneously and can be computed using minimal network data including (1) bus connectivity, (2) line susceptances, (3) line loading limits, and (4) bus power injection limits. The metrics provide an overall measure of grid inadequacy that is not dependent on a given operating condition. The first set of metrics generalizes the sampling-based approach in [8] and [9] under the presented framework. As such, the first metric is very much dependent on the quality of the sample and can be blind outside of the identified states. The other two sets of metrics are original and does not have this dependency. The underlying assessment against the whole power injection space fits well with an ideal future grid that provides non-discriminatory access to all participants that accommodates a diversity of power exchanges. Moreover, we decouple contingency analysis and power flow control interventions in the operations modeling in order to highlight the inherent properties of the network. However, as with other metrics, the presented metrics are only as good as the models and assumptions used. We will comment on these and identify potential applications.

## III. PROPOSED FRAMEWORK TO MEASURE GRID INADEQUACY

### A. Infeasible power system operation

Power grids facilitate the exchange of electrical energy between generators and loads through interconnecting transmission lines. Feasible power exchanges (in the static, non-contingency, and non-intervention sense) must satisfy not only strict physical laws governing the power flows but also equipment and operational constraints including bus power injection, line power flow, and voltage magnitude constraints. A large infeasible power injection set means that constraints are violated for many power exchange schedules and is indicative of the need to deploy preventive of corrective measures.

If possible, a network with a small infeasible power injection set is preferred because this means that the network can accommodate diverse power injection scenarios on its own even without preventive or corrective interventions. However, characterizing the infeasible set using the non-linear AC power flow model is generally complicated and computationally intensive [12]. The DC power flow model is a linear approximation of the full AC model [13] and makes the analysis of feasible and infeasible sets simpler at the expense of model fidelity and accuracy. This model is ubiquitous in research and practice, in both the operations and infrastructure planning applications, evident in the literature reviewed in Section II.

### B. Infeasible set of the constrained DC power flow model

In the DC power flow model, infeasible operation is mainly characterized by line overloading. That is, a vector of bus power injections $\boldsymbol{P^s} = [P_1^s, P_2^s, \dots, P_n^s]^{\mathrm{T}}$ representing scenario $s$ is infeasible in grid $\mathcal{g}$ with $l$ lines and $n$ buses if the following relations are satisfied except (1):

$$|f_{ab}^{s,\mathcal{g}}| \leq C_{ab}^{\mathcal{g}} \qquad \forall ab \in \mathcal{E}^{\mathcal{g}} \qquad (1)$$

$$f_{ab}^{s,\mathcal{g}} = B_{ab}^{\mathcal{g}}\left(\theta_a^{s,\mathcal{g}} - \theta_b^{s,\mathcal{g}}\right) \qquad \forall ab \in \mathcal{E}^{\mathcal{g}} \qquad (2)$$

$$P_i^s = \sum_{j \in \mathcal{V}^{\mathcal{g}}, j \neq i} f_{ij}^{s,\mathcal{g}} \qquad \forall i \in \mathcal{V}^{\mathcal{g}} \qquad (3)$$

$$\sum_{i \in \mathcal{V}^{\mathcal{g}}} P_i^s = 0 \qquad (4)$$

$$P_i^{min} \leq P_i^s \leq P_i^{max} \qquad \forall i \in \mathcal{V}^{\mathcal{g}} \qquad (5)$$

where $f_{ab}$ is the power flow and $C_{ab}$ is the capacity of line $ab$ with end buses $a$ and $b$, $\mathcal{E}$ is the set of lines with cardinality $l$, $B_{ab}$ is the susceptance of line $ab$, $\theta_a$ is the voltage angle at bus $a$, $P_i$ is the power injection at bus $i$, $\mathcal{V}$ is the set of buses labelled from 1 to $n$, $P_i^{min}$ and $P_i^{max}$ are the power injection limits at bus $i$, and the superscripts indicate dependence in $s$ and $\mathcal{g}$. Eq. (1) limits the power flow in a line within its capacity. Eqs. (2) and (3) are the DC power flow analog of Ohm's law and Kirchhoff's Current Law (KCL) respectively. Eq. (4) is the power balance equation and (5) keeps power injection variability within limits.

Depending on the type of analysis, the power injection limits can represent various sources of uncertainty. They can capture long-term uncertainty in spatiotemporal load and generation growth as well as short-term uncertainty in market dispatch, load flexibility and variability, energy storage charge and discharge capabilities, and renewable energy intermittency. Detailed models can capture the interrelation between bus power injections and limit the spanned power injection space if data is available, observed trends are deemed to hold in the planning horizon, or market simulation tools are deemed reliable. Otherwise, interval sets provide a viable alternative in addition to giving information on infeasible operation that are not captured by historical or market-based models.

### C. Framework to measure grid inadequacy

Any power injection vector satisfying (4)-(5) represents a valid power exchange. However, some vectors can be infeasible due to network congestion, i.e. when at least one line is overloaded. Power flow control via topology, susceptance, or voltage angle control or power injection control via generation rescheduling or load curtailment can mitigate some congestion scenarios. The prevalence of grid congestion and the need to use preventive or corrective control strategies highlight inherent grid inadequacy to accommodate power injection diversity on its own without intervention.

A way to quantify this inherent inadequacy is to compare the size of the following sets:

- Set $\mathcal{F}$ of valid power injections satisfying (4)-(5) and
- Set $\mathcal{I}$ of infeasible power injections satisfying (2)-(5) but fail to satisfy (1) for at least one line.

The larger the size of Set $\mathcal{I}$ relative to Set $\mathcal{F}$, the more inadequate a grid inherently is. However, solving for the size of these sets is complicated especially for large $n$ and $l$. Interpreting the constrained DC power flow model as a vector mapping from one space to another and projecting the resulting surfaces in lower dimension can circumvent this challenge.

*D. Visualizing the constrained DC power flow model*

The constrained DC power flow model can also be formulated using matrix operations as derived in detail in [14]. That is, a power injection scenario is feasible for a given grid if the following relations are met including (4):

$$\left| \boldsymbol{f}^{s,\mathscr{g}} \right| \leq \boldsymbol{C}^{\mathscr{g}} \tag{6}$$

$$\boldsymbol{f}^{s,\mathscr{g}} = \boldsymbol{X}^{\mathscr{g}} \boldsymbol{K}^{\mathscr{g}} \boldsymbol{\theta}^{s,\mathscr{g}} \tag{7}$$

$$\boldsymbol{P}^{s} = \boldsymbol{B}^{\mathscr{g}} \boldsymbol{\theta}^{s,\mathscr{g}} \tag{8}$$

$$\boldsymbol{P}^{min} \leq \boldsymbol{P}^{s} \leq \boldsymbol{P}^{max} \tag{9}$$

where $\boldsymbol{f}$ is the vector of line power flows, $\boldsymbol{C}$ is the vector of line capacities, $\boldsymbol{X}$ is the diagonal matrix of line susceptances, $\boldsymbol{K}$ is the edge-to-bus incidence matrix, $\boldsymbol{\theta}$ is the vector of bus voltage angles, $\boldsymbol{P}$ is the vector of bus power injections, $\boldsymbol{B}$ is the full DC power flow susceptance matrix, $\boldsymbol{P}^{min}$ and $\boldsymbol{P}^{max}$ are the vectors of bus power injection limits, and the superscripts denote dependence in $s$ and $\mathscr{g}$. Eqs. (6) and (9) are the matrix representation of (1) and (5) respectively while (7) and (8) are the matrix equivalent of (2) and (3) respectively. We can combine (7) and (8) to eliminate $\boldsymbol{\theta}$ as follows:

$$\boldsymbol{f}^{s,\mathscr{g}} = \boldsymbol{X}^{\mathscr{g}} \boldsymbol{K}^{\mathscr{g}} (\boldsymbol{B}^{\mathscr{g}})^{+} \boldsymbol{P}^{s} \tag{10}$$

where $(\boldsymbol{B})^{+}$ is the pseudoinverse of $\boldsymbol{B}$. Eq. (10) highlights the cause-and-effect relationship between $\boldsymbol{P}$ and $\boldsymbol{f}$ in terms of the grid topology and network parameters. This also allows us to conveniently interpret the constrained DC power flow model as a vector mapping as follows.

Eq. (10) can be interpreted as a mapping of vector $\boldsymbol{P}$ in the $n$-dimensional power injection space, *P-space*, to the $l$-dimensional line power flow space, *f-space*, through the linear mapping function $\boldsymbol{XKB}^{+}$. This becomes a constrained mapping when combined with Eqs. (4), (6), and (9). Fig. 2 visualizes this mapping for a three-bus three-line grid. The susceptances are 1.0 p.u. for Line 1-2 and 0.5 p.u. for Lines 1-3 and 2-3. The power injection limits are 0.0–2.0 p.u. for Bus 1 and -1.0–0.0 p.u. for Buses 2 and 3. The line capacities are all 1.0 p.u.

In Fig. 2, an $n$-dimensional hypercube defined by (9) bounds all possible power injection combinations in *P-space*. A hypersurface defined by (4) lies within the hypercube and characterizes Set $\mathcal{F}$ of valid power exchanges. Linear mapping using (10) projects this hypersurface into *f-space*. Portions of the projection fall inside or outside an $l$-dimensional bounding hypercube defined by (6) in *f-space*. Power exchanges projected outside the bounding hypercube have one or more congested lines and hence are infeasible. This characterizes Set $\mathcal{I}$. The larger the projection outside the *f-space* bounding hypercube, the larger the size of the infeasible power injection set and the more inadequate a grid inherently is to accommodate power injection diversity on its own.

Traditional methods in designing adequate grids ensure that the point representing the peak load dispatch or points representing a number of power injection scenarios lie inside the *f-space* bounding hypercube, with or without employing power flow or power injection control. However, increased power injection diversity and increased demand for nondiscriminatory access in future grids may require a paradigm shift. It will become increasingly important not just to accommodate a number of power injection points but instead accommodate a larger area of the power injection hypersurface as economically as possible. The ideas in this paper help motivate this paradigm shift and will be useful future developments in this field.

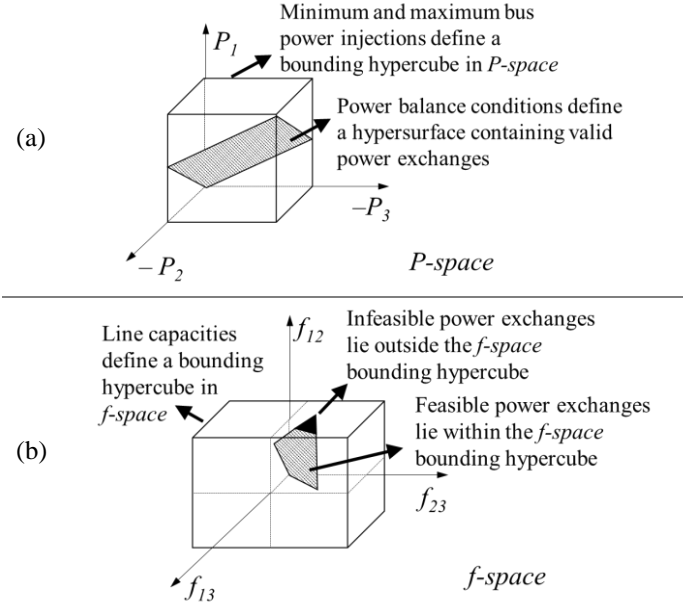

Fig. 2: Visualization of the constrained DC power flow model as a vector mapping: (a) a hypersurface represents the set of valid power injection scenarios in the power injection space, *P-space*, and (b) the vector projection of the hypersurface within the bounding hypercube in the line power flow space, *f-space*, determine feasibility.

*E. Visualizing the f-space surfaces in 1D*

Usually, projections of feasible spaces are made in *P-space* as was done in [2]-[7], [10]. This is generally complicated because the feasible range of power injection on a bus is dependent on the power injections at other buses. We present below a complementary approach that projects instead in *f-space*, a non-standard practice as far as the authors are aware. This is relatively much less convoluted and provides new insights on grid vulnerabilities and bottlenecks by identifying lines that are prone to congestion when power injections are not adequately scheduled. We return to the problem of projecting in *P-space* in the next section.

We can project the feasible and infeasible spaces in *f-space* along the *f-space* basis vectors to be able to visualize and analyze the high-dimensional surfaces involved. The projection of the feasible space can be obtained by solving for the maximum feasible line flows for each line in both directions while considering all line capacity limits. The projection of the whole power injection hypersurface can be obtained by solving for the maximum line flows for each line in both directions while ignoring all line capacity limits. Superimposing these two sets of parameters will show the extent of the infeasible space, projected in terms of individual line power flows. The following mathematical models can be used to solve these parameters.

The projection of the power injection hypersurface is spanned by the maximum unconstrained line flows in a line in both directions, $f_{\max\_u,ab}$ and $f_{\max\_u,ba}$, with all line capacity constraints ignored. The difference determines a line's unconstrained loading range. The following linear program in $\boldsymbol{\theta}$, $\boldsymbol{P}$, and $f$ solves for $f_{\max\_u,ab}$. A similar linear program can be used to solve for $f_{\max\_u,ba}$.

Objective $\qquad f_{\max\_u,ab} = \max f_{ab}$ (11a)

Subject to $\qquad$ Eqs. (2)-(5) (11b)

On the other hand, the projection of the feasible subspace of the power injection hypersurface is spanned by the maximum feasible line flows in a line in both directions, $f_{\max\_f,ab}$ and $f_{\max\_f,ba}$, with all line capacity constraints considered. The difference determines a line's feasible loading range. The following linear program, along with a similar program, can be used to solve for these parameters.

Objective $\qquad f_{\max\_f,ab} = \max f_{ab}$ (12a)

Subject to $\qquad$ Eqs. (1)-(5) (12b)

After solving for the maximum unconstrained and feasible line flows, we can superimpose the loading ranges and line capacities in a line loading range diagram. Fig. 3 shows this for the three-bus network in Section II-D. Red outlines represent the line capacities. White areas represent excess line capacity and remains unused unless the power injection limits change, the network is modified by design or by accident, or non-injection power flow control strategies are implemented. The black areas identify infeasible operation.

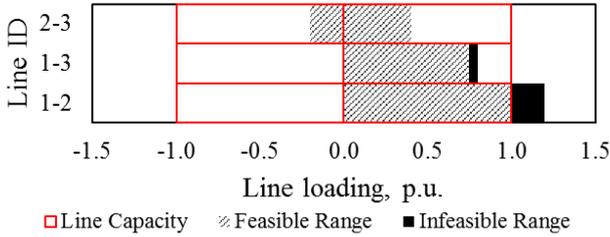

Fig. 3: A line loading range diagram showing the one-dimensional visualization of the *f-space* bounding hypercube (outlined areas) and hypersurface (shaded areas) for the three-bus three-line network. The presence of black areas outside the outlined areas indicate infeasible operating states.

The presence of infeasible areas does not necessarily mean that the network cannot satisfy peak or maximum loading conditions. Rather, the infeasible areas indicate that some loading states served with a given set of generator dispatch, which may or may not correspond to system peak conditions, may overload a given line. There may or may not exist a set of dispatch able to serve peak demand. As such, the proposed procedure in this section is not meant to measure adequacy to meet peak demand but rather to highlight grid inability to serve a diversity of dispatch scenarios. The procedures in the next section can answer the question of adequacy during peak loading and other loading conditions.

In the three-bus example, there is ample capacity in Lines 1-3 and 2-3 to cover all possible diversity of power injection scenarios within bus power injection limits. However, capacity constraints in Line 1-2 prohibit the dispatch of some power injection scenarios. The prevalence of the infeasible projections is indicative of network inability to accommodate power injection diversity and the need to deploy preventive or corrective interventions. Moreover, the infeasible areas indicated by this procedure can be used to motivate more detailed analyses. For example, the families of power injection scenarios that lead to infeasible states and the likelihood of these states occurring merit further exploration. Since it is the first time that this approach is presented in the context of grid inadequacy assessment, the authors hope that future work will find other suitable uses for it.

### F. Visualizing the P-space surfaces in 1D

As we mentioned in Section III-E, projecting the *P-space* surfaces to a lower dimension to show the infeasible set is challenging. Previous works such as [2]-[7], [10] have tried various approaches along this direction and have their own set of capabilities and limitations. One limitation is the dependence on an operating point. This is as expected because the feasible range of power injection in one bus is dependent on the power injections in other buses. In this section, we present an approach that projects in a space related to *P-space*. The choice of a different space removes the complications of dependence on individual bus power injection. This is not meant to replace existing methods however and is intended to provide a complementary approach that provides useful information in a different form.

The approach we use is to project along the total network loading vector, $L_t$, given by the sum of the loads at each bus. The idea is to find thresholds of total network loading that trigger congestion instead of finding allowable power injection ranges as is done in [2]-[7], [10]. Under the DC power flow model assumptions, we can find $L_{t-}$ and $L_{t+}$ such that $L_{t-}$ is a network loading threshold below which congestion is not possible regardless of the generation schedule and $L_{t+}$ is the maximum network loading that the grid can accommodate without congestion through at least one generation schedule. We can solve for $L_{t-}$ in two steps: (1) solving for $L_{t,ab}$, the minimum load that triggers congestion of line $ab$ in direction $ab$ and (2) solving for the minimum of the values obtained in Step 1 for all lines in both the forward and reverse power flow directions. That is,

$$L_{t-} = \min\{L_{t-,ab} \; \forall ab \in \mathcal{E}^*\} \qquad (13a)$$

where $\mathcal{E}^*$ is the set of line indices for both forward and reverse power flow directions. Here, each $L_{t-,ab}$ is calculated by solving the following linear program:

Objective $\qquad L_{t-,ab} = \min L_t$ (13b)

Subject to $\qquad f_{ab} \geq \mathcal{C}_{ab}$ (13c)

$\qquad$ Eqs. (2)-(5) (13d)

Likewise, we can solve for $L_{t+}$ as follows:

Objective $\qquad L_{t+} = \max L_t$ (14a)

Subject to $\qquad$ Eqs. (1)-(5) (14b)

After solving for the relevant points, we can make the one-dimensional visualization in Fig. 4 where we can also superimpose the maximum total load, $L_{max}$. This diagram visualizes the network loading ranges that are congestion-free wherein $L_t \leq L_{t-}$ and congestion-prone wherein $L_t > L_{t-}$. A subset of the latter set is congestion-positive for $L_t > L_{t+}$ if $L_{t+} < L_{max}$.

This approach provides grid adequacy and inadequacy information in terms of congestion-free, congestion-prone, and congestion-positive network loading ranges. For the three-bus example, operating below $L_{t-} = 1.5$ p.u. guarantees no congestion while operating above $L_{t+} = 1.75$ p.u. guarantees congestion regardless of generation schedule under DC power flow model assumptions. Since $L_{max} = 2.0$ p.u. exceeds $L_{t+}$, there is no feasible generation schedule for the maximum load. These information can motivate further studies that explore the cause of these thresholds as well as help quantify inherent grid inadequacy to serve loads. For example, a low value of $L_{t-}$ relative to $L_{max}$ is indicative of (1) a grid that needs a lot of operational intervention to serve demand or (2) a selective grid that only allows non-discriminatory access up to a certain loading level.

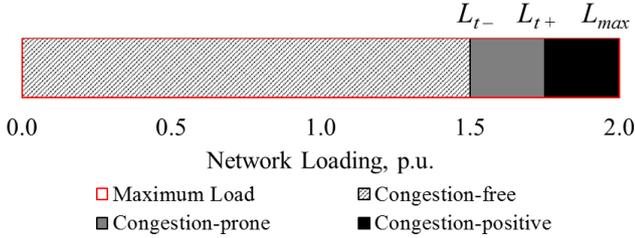

Fig. 4: One-dimensional visualization of the *P-space* surfaces using the total network loading.

## IV. PROPOSED MEASURES OF GRID INADEQUACY

This section extends the framework developed in Section III to define metrics of grid inadequacy. We present one sampling-based metric and two metrics derived from the projection of the feasible and infeasible sets in lower dimension.

### A. Scenario Pool Infeasibility Ratio (SPIR)

We can define a representative scenario pool $\Omega_S$ of $T$ power injection vectors $\boldsymbol{P}^{s_1}, \boldsymbol{P}^{s_2}, ..., \boldsymbol{P}^{s_T}$ to represent the set of valid power exchanges. Each vector represents a power injection scenario that satisfy (4) and (5). System state selection requires a balance between adequate space representation and computational tractability. It is a challenging problem on its own and hence is placed outside the scope of this work. Some of the common approaches include the following. If data is available and trends are assumed to hold in the planning horizon, probability distribution and correlation models may be used to generate scenarios. Market simulation can also be used if there are reliable datasets and models. Otherwise, combinatoric sampling or worst-case sampling become viable options.

After defining the scenario pool, we can evaluate each scenario using DC power flow for a given network topology $\mathscr{g}$ and solve for the number of line overloading instances per scenario $N_{Ls}$, the number of line overloading instances in the scenario pool $N_L$, and the number of scenarios with congestion $N_S$ as follows:

$$N_{Ls}^{s,\mathscr{g}} = \sum_{ab \in \mathcal{E}^{\mathscr{g}}} \left\{ u\left(f_{ab}^{s,\mathscr{g}} - C_{ab}^{\mathscr{g}}\right) + u\left(-f_{ab}^{s,\mathscr{g}} - C_{ab}^{\mathscr{g}}\right) \right\} \quad (15)$$

$$N_L^{\Omega_S,\mathscr{g}} = \sum_{s \in \Omega_S} N_{Ls}^{s,\mathscr{g}} \quad (16)$$

$$N_S^{\Omega_S,\mathscr{g}} = \sum_{s \in \Omega_S} u\left(N_{Ls}^{s,\mathscr{g}}\right) \quad (17)$$

where $u(x)$ is a function that returns 1 if $x > 0$ and returns zero otherwise and the superscripts show the dependence on $s$, $\mathscr{g}$, and $\Omega_S$. Using $N_S$ and $T$, we can then define the *Scenario Pool Infeasibility Ratio* (*SPIR*) for a given grid and scenario pool as follows:

$$SPIR^{\Omega_S,\mathscr{g}} = \frac{N_S^{\Omega_S,\mathscr{g}}}{T^{\Omega_S}} \quad (18)$$

This metric captures the proportion of power exchanges in the scenario pool that is infeasible. Using the sets defined in Section III-C, we use $N_S$ to measure Set $\mathcal{I}$ and $T$ to measure Set $\mathcal{F}$. The more representative the scenario pool is of the actual *P-space* hypersurface, the better the approximation given by *SPIR* of the relative sizes of Sets $\mathcal{F}$ and $\mathcal{I}$.

If the assessment is made using the same scenario pool and between grid topologies with the same buses, we can use $N_S$ to compare grid alternatives directly. Furthermore, since $N_L$ is related to $N_S$ as implied in (15)-(17), $N_L$ is also useful to compare options. Dividing $N_L$ with the number of line flow assessments given by $T \cdot l$ gives the *Scenario Pool Overloading Ratio* (*SPLOR*) as follows:

$$SPLOR^{\Omega_S,\mathscr{g}} = \frac{N_L^{\Omega_S,\mathscr{g}}}{T^{\Omega_S} \cdot l^{\mathscr{g}}} \quad (19)$$

*SPIR* is a generalization of the ad hoc approaches in [8] and [9] to the whole power injection space. It shares the same underlying framework with the *ENLC* metric from [1] but counts grid congestion states instead of load curtailment states under the grid inadequacy assessment framework outlined in Section III. As such, it is a flexible metric such that power flow control and contingency analysis can also be integrated in the computation by replacing the basic DC power flow model with more appropriate model. However, as with all other sampling-based metrics, *SPIR* is highly dependent on the quality of the selected states and is potentially blind to other states outside this sample. The next two metrics do not require the identification of discrete system states and can be used to complement *SPIR*.

### B. Congestion-prone Network Loading Ratio (CPNLR)

Using the notations in Section III-F, we can use $L_{t-}$ and $L_{max}$ to define the *Congestion-free Network Loading Ratio* (*CFNLR*) as follows:

$$CFNLR^{\mathscr{g}} = \frac{L_{t-}^{\mathscr{g}}}{L_{max}} \quad (20)$$

This metric gives the proportion of the maximum network loading that is always free from congestion under the DC power flow model assumptions. Similarly, the ratio of $L_{max} - L_{t-}$ and $L_{max}$ gives the *Congestion-prone Network Loading Ratio* (*CPNLR*) as follows:

$$CPNLR^{\mathscr{g}} = 1 - CFNLR^{\mathscr{g}} = \frac{L_{max} - L_{t-}^{\mathscr{g}}}{L_{max}} \quad (21)$$

This metric gives the proportion of maximum network loading that may result in congestion if the power injections are not properly scheduled. Using the sets defined in Section III-C, we use $L_{max}$ to give a measure for Set $\mathcal{F}$ and $L_{max} - L_{t-}$ for Set $\mathcal{I}$ in *P-space*.

A limitation of *CPNLR* is that it overestimates grid inadequacy while *CFNLR* underestimates adequacy. Since operating the grid beyond $L_{t-}$ may or may not result in network congestion depending on the chosen generation schedule, $L_{max} - L_{t-}$ gives an inflated measure of Set $\mathcal{I}$. As such, *CPNLR* also gives an inflated measure of grid inadequacy. Regardless, *CPNLR* is still useful in revealing the extent of grid neediness for careful monitoring and potential interventions as a result of not being able to serve power exchanges on its own. An ideal value of zero means that all loading levels can be served by any combination of generator dispatches without necessary intervention. However, this only guarantees adequacy from a non-contingency perspective under DC power flow assumptions. It must be used with its limitations in mind and only in an appropriate context with other decision criteria considered.

### C. Total Line Overloading Ratio (TLOR)

The visualization in *f-space* in Section III-E estimates the extent of infeasible operating ranges of each line given the variability in power injection in the buses. From the visualization in Fig. 3, we can define the *Infeasible Line Loading Margin (ILLM)*, *Unconstrained Line Loading Range (ULLR)*, and *Infeasible Line Loading Ratio (ILLR)* for each line as follows:

$$ILLM_{ab}^{\phi} = f_{\max\_u,ab}^{\phi} - f_{\max\_f,ab}^{\phi} + f_{\max\_f,ba}^{\phi} - f_{\max\_u,ba}^{\phi} \quad (22)$$

$$ULLR_{ab}^{\phi} = f_{\max\_u,ab}^{\phi} - f_{\max\_u,ba}^{\phi} \quad (23)$$

$$ILLR_{ab}^{\phi} = \frac{ILLM_{ab}^{\phi}}{ULLR_{ab}^{\phi}} \quad (24)$$

For a given line $ab$, $ILLM_{ab}$ gives the difference between the maximum unconstrained line flow and feasible line flow, $ULLR_{ab}$ gives the difference between the maximum unconstrained line flows in both directions, and $ILLR_{ab}$ gives the ratio of the two.

We can then define the *Total Infeasible Line Loading Ratio (TILLR)* as a measure of overall grid inadequacy. It uses the *ILLM* and *ULLR* of all the lines to compute a metric like *ILLR* for the whole system as follows:

$$TILLR_{ab}^{\phi} = \frac{\displaystyle\sum_{ab\in\mathcal{E}^{\phi}} ILLM_{ab}^{\phi}}{\displaystyle\sum_{ab\in\mathcal{E}^{\phi}} ULLR_{ab}^{\phi}} \quad (25)$$

Using the sets defined in Section III-E, we use the total *ULLR* as a measure of Set $\mathcal{F}$ and the total *ILLM* as a measure of Set $\mathcal{I}$. As with *CPNLR*, this representation is not exact because of the limitations in the low-dimensional projections and the redundancy in accounting for the projected spaces. Nevertheless, *TILLR* gives a measure of grid inadequacy that can be useful indicators of the extent of network vulnerabilities.

The individual *ILLRs* give the proportion of the infeasible loading range relative to the total unconstrained loading range of a line. This can be used as an indicator of line adequacy if the value is zero or an indicator for overloading and network bottleneck otherwise. While a similar physical interpretation cannot be made for *TILLR*, it can be seen as the extent of additional power injection diversity that the grid can accommodate only if there are no line capacity constraints. As with *CPNLR*, the ideal value of *TILLR* is zero. If this is the case, no lines in the network will ever get congested even without intervention if power injection limits are satisfied. This guarantee, likewise, is limited to the non-contingency case under DC power flow assumptions. Additional checks are required to assess adequacy and inadequacy outside the scope of the framework presented in this paper.

## V. Illustrative Examples

### A. 6-bus test system

The relevant data for the 6-bus test system is available in [15]. Of the six buses, three have generators and five have loads. There are a total of thirteen installed lines.

To construct the scenario pool for *SPIR* evaluation, we use a combinatoric sampling approach since it is tractable for this test system. We sample $T = 3,139,831$ power injection scenarios that represent eleven negative power injection levels at 0%, 10%, …, 100% of maximum and eleven positive power injection levels at 0%, 5%, 15%, …, 95% of maximum ±5% to balance out the demand. The scenarios represent power injections from traditional loads and generators as well as power injections from renewables, dynamic loads, and energy storage. We specifically chose not to exceed the 100% threshold to illustrate the potential for network congestion even if power injections remain within limits.

Fig. 5 visualizes the distribution of the six-dimensional power injection scenario pool in reduced 3D space obtained using principal component analysis (PCA) [15]. The axes are the top three principal components. Points corresponding to the scenarios with and without congestion are interspersed with one another, making it difficult to gauge the relative sizes of the infeasible and feasible sets. This challenge reinforces the utility of the proposed inadequacy metrics in quantifying the relative sizes of these sets.

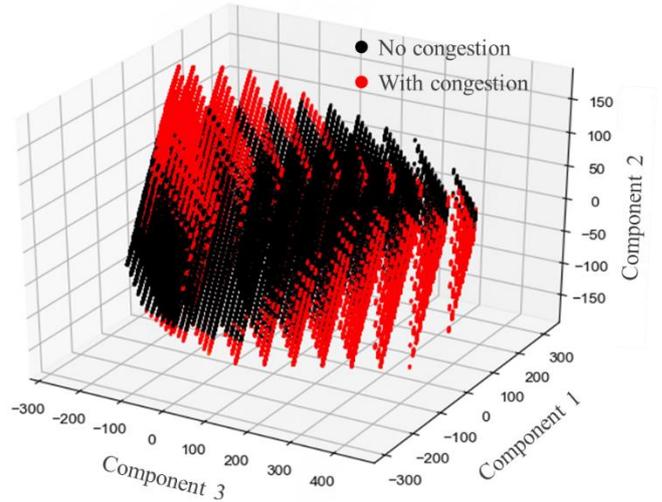

Fig. 5: Three-dimensional visualization of the 6-bus test system infeasible set using a scenario pool ($T = 3,139,831$) and PCA.

*SPIR* is equal to 33.1% implying that a third of the sampled scenarios resulted in network congestion. The infeasible scenarios can be analyzed further to determine the family of power injection scenarios that congest the network, explore the likelihood of such scenarios, and then devise appropriate pre-ventive or corrective interventions. *CPNLR* is equal to 82.1% with $L_{t-}$ = 135.7 MW and $L_{max}$ = $L_{t+}$ = 760 MW. This means that operating the grid above $L_t$ = 135.7 MW has potential for congestion and requires coordinated scheduling of bus power injections. Since $L_{max} = P_{t+}$, there is at least one generation schedule that can serve the maximum total load. In a way, the grid is "needy" since a large proportion of network loading requires careful monitoring and potential operational intervention. Likewise, without adequate power flow control interventions, the grid can also be seen as inherently selective from a market perspective because it discriminates against some schedules in serving a particular loading level.

Finally, *TILLR* is equal to 27.1%. Fig. 6 shows how each line contributes to this value. The diagram shows that only Line 2-4 will never become overloaded regardless of the power injection distribution. The diagram also identifies lines that limit a grid's ability to accommodate diverse power injection states. This can serve as basis for later analyses on what kinds of power injection distributions are affected by grid bottlenecks and what remedial action can be made.

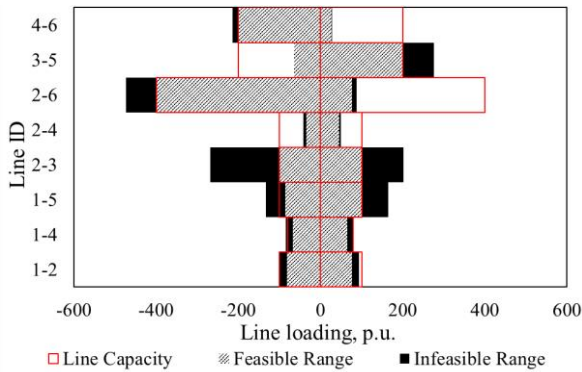

Fig. 6: Line loading range diagram of the 6-bus test system.

### B. 118-bus test system

We perform a similar analyses using the 118-bus test system using the data in [16]. The grid has 118 buses and 186 lines, with generation at 54 buses and loads at 91 buses.

Since defining discrete load and generation levels to define the scenario pool is computationally prohibitive, we sample from the line loading space, including low and intermediate loading scenarios as well as worst case scenarios that result in maximum line loading. We define ten loading levels for each line by dividing each line's unconstrained loading range into equal parts. We then solve for the power injection scenarios that result in each line loading level $F_{ab}$ at minimum network loading as follows:

Objective $$\min L_t \tag{26a}$$

Subject to $$f_{ab} = F_{ab} \tag{26b}$$

$$\text{Eqs. (2)-(5)} \tag{26c}$$

This process yielded $T = 1,860$ scenarios that represent a spectrum of network loading conditions.

Around forty percent of the sampled scenarios result in congestion with *SPIR* = 43.4%. *CPNLR* is equal to 94% with $L_{t-}$ = 222.9 MW and $L_{max}$ = $L_{t+}$ = 3,733 MW. A low value of $L_{t-}$ relative to $L_{max}$ indicates that congestion is expected even at very low loading levels if there is no coordinated generation scheduling. This is indicative of the grid's inherent selectivity in serving demand unless power flow control options are available. Since $L_t = L_{t+}$, there should be at least one feasible generation schedule able to meet the maximum demand. However, this schedule may not necessarily have the least cost or the most renewable energy utilization rate. *TILLR* is equal to 31.5%. Fig. 7 shows the top ten lines with the largest *ILLM*s.

These results provide indicators of inherent grid inadequacies, vulnerabilities, and bottlenecks that merit further exploration and the subsequent design of preventive and corrective interventions, if deemed necessary.

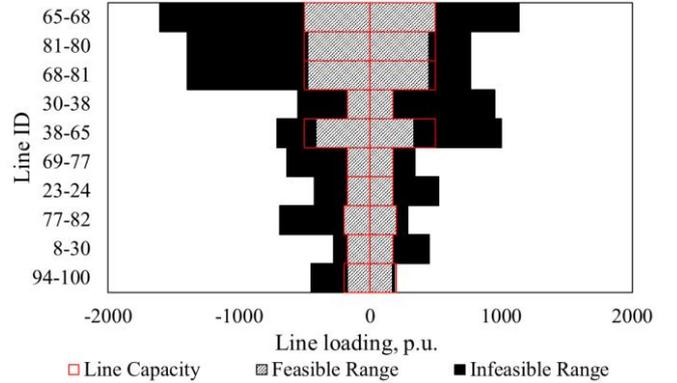

Fig. 7: Line loading range diagram of the 118-bus test system showing the top ten lines with the largest *ILLM*.

## VI. REMARKS, FUTURE WORK, AND POTENTIAL USE-CASES

As we discussed in Section II-A, grid adequacy assessment requires assumptions on network and operations modeling and system state selection. The level of fidelity of the models used and the comprehensiveness of system states represented depends on the accuracy and tractability requirements in the desired use-case. The proposed approach chooses a specific set of assumptions that defines its capabilities and limitations. We comment on these in this section and identify interesting future work and potential use cases.

### A. Remarks on grid and operations modeling assumptions

In the proposed framework, the DC power flow model is used to assess grid inadequacy in a non-contingency and strictly non-intervention setting. As with the other works reviewed that uses the DC power flow model, the results are subject to the model's inherent limitations. Despite its limitations, however, this model remains widely-used in recent work such as in [7] in either operations planning or infrastructure planning setting. It will be interesting future work to look at counterparts to the metrics proposed using other power flow models and then compare the results.

Likewise, using a non-contingency and non-intervention approach means that the resulting inadequacy metrics does not reflect the full capabilities of a grid to manage power exchanges in both normal and contingency conditions. What the metrics provide are indicators of *inherent* grid inadequacy, vulnerabilities, and bottlenecks to accommodate power injection diversity on its own. These indicators are useful in gauging the extent of necessary intervention the grid requires and the inherent readiness of a grid for non-discriminatory access. As such, the proposed metrics are not meant to replace other adequacy and inadequacy metrics but rather provide comple-mentary information. Adequacy and inadequacy assessment considering operational interventions and contingency scenarios will still be necessary. Future work can look at ways to incorporate operational interventions and contingency modeling in the proposed framework. Possible approaches for the latter include deterministic N-1 analysis or probabilistic risk assessment.

### B. Remarks on system state selection approach

In terms of system state selection, the proposed metrics uses two approaches. The first metric uses a sampling-based approach to identify discrete system states. This is similar to the approaches used in [1], [8], and [9] but generalized to the whole power injection space. As with other approaches that use this framework, it is a challenge to identify systems states that best represent future states while keeping the size tractable. Scenario selection and scenario reduction methods can be used for these types of cases and will remain relevant areas of research in the future. However, planners must remember that such an approach can be blind to system states not included in the samples. As such, additional analysis must be made to explore these blind spots, assess their likelihood, and design appropriate interventions. The approach used in the two other metrics helps in this regard.

The other two metrics implicitly considers all system states using interval sets that represent the full power injection space. The approach used is similar to those in [7] and [11] using loadability and feasibility sets but further developed into tractable metrics tailored specifically for inadequacy assessment. While this approach is conservative, it complements sampling-based approaches by providing insights on potential blind spots not captured in sampling-based approaches. Using interval sets also becomes particularly useful when historical power injection distribution data is not readily available, historical trends are deemed inapplicable in the planning horizon, market simulation models are deemed unreliable, or when planners would like to explore grid inadequacy against previously unobserved trends. These conditions are especially applicable in an expansion planning setting with increased uncertainty that we comment on next.

### C. Potential use cases

Adequacy assessment using sampling-based metrics such as *SPIR* are very flexible and can reflect use-case requirements by changing the network and operation model used as well as the system state selection process. As such, it can easily find potential applications in adequacy assessment in either operations planning or expansion planning provided that the appropriate set of assumptions are applied.

The other two metrics, *CPNLR* and *TILLR*, are more specialized because of the underlying models used. These metrics tests for inadequacy against all possible power injection scenarios which may be impractical in some cases. In the operations setting for example, hour-ahead and day-ahead system states are much more likely to resemble previously observed system states from the recent past. As such, the level of variability of future states against which adequacy or inadequacy must be assessed is much more limited. In this case, adequacy assessment against detailed distribution and correlation models, market simulation models, or tighter interval sets are more appropriate. Testing against potential blind spots is still neces-sary but not in the scale of exploring the whole power injection space.

On the other hand, power injections years or decades ahead in an expansion planning setting are much more uncertain and diverse. This is especially true today when data and models on how future market players owning one or a combination of renewable generation, dynamic loads, and energy storage are not yet available. In this application, the assumptions underlying the *CPNLR* and *TILLR* metrics become viable preliminary options until more detailed data and simulation models become available. Even if detailed models become available, these can have blind spots and it will still be necessary to scan the whole power injection space for potential vulnerabilities. Since no model will be accurate enough to predict the long-term future with certainty, information on potential grid vulnerabilities that the proposed framework and metrics provide will allow planners to see the network's limitations and subsequently devise appropriate interventions to hedge against likely problematic future scenarios.

In this regard, future work can look at the development of decision support tools that use the metrics presented in identifying grid expansion designs that are robust against power injection diversity in future grids. A suitable application is in initial stages of expansion planning where a small subset of interesting grid expansion plans needs to be identified from a large combinatoric pool. One approach is to use a greedy heuristic where the metrics are minimized by adding one line at a time. More sophisticated approaches would involve optimization using mathematical or metaheuristic models that try to find local or global minima. In such applications, other screening tools can be used, each with its own set of capabilities and limitations, to generate a subset of interesting grid designs. The shortlisted options can then be subject to more rigorous multi-criteria assessment including PQ balancing studies, congestion management, contingency response, cost-benefit analysis, market simulations, stability analysis, and robustness assessment against different realizations of uncertain parameters. And as new information becomes available, existing plans can be reassessed or the whole process repeated altogether.

## VII. Conclusion

The integration of more intermittent renewable sources, dynamic loads, and energy storage on top of a competitive market environment will require a non-discriminatory grid that can accommodate diverse power injection states. This work presents a new framework and set of metrics that measure grid inadequacy to accommodate increasing levels of power injection diversity. Such a framework and set of metrics comprise fundamental building blocks in assessing and designing future grids.

The framework uses the infeasible set of the constrained DC power flow model to quantify grid inadequacy against power injection diversity. Since it is complicated to compute the actual size of the high-dimensional sets, we present three approaches to give a measure to the sets involved. One uses a sampling-based approach while the other two uses novel dimension reduction techniques. We then use these ideas to formulate three inadequacy metrics that describe different but complementary grid inadequacy properties. Illustrative exam-ples using the 6- and 118-bus test systems show that the metrics can be used as indicators of grid readiness to accommodate diverse operating states as well as indicators of grid vulnerabilities and bottlenecks.

Like other metrics, the proposed metrics capture a limited aspect of grid inadequacy assessment defined by the network and operations model used as well as the system state selection process chosen. We discuss the implications of the choice of assumptions made and made a case for the use of the proposed framework and metrics in an expansion planning setting. We identified interesting areas for future work including the use of more detailed power flow models in developing similar metrics, appropriate scenario selection and scenario reduction techniques to balance space representation and model tractability, integration of operational interventions and contingency analysis in the proposed framework, and development of decision-support tools for identifying candidate grid expansion plans.

We close by noting that the problem of grid adequacy assessment considering power injection diversity against more intermittent renewable generation, dynamic loads, and energy storage is timely and important. The framework and metrics presented provide a new and different approach to assessing grid inadequacy that is potentially useful in future research and practice. We identified a lot of interesting areas for future work that we hope can inspire other researchers to study this problem.

## REFERENCES

[1] R. Billinton and W. Li, *Reliability Assessment of Electric Power Systems Using Monte Carlo Methods*. New York: Springer Science + Business Media, 1994.

[2] J. Zhu, R. Fan, G. Xu, and C. S. Chang, "Construction of maximal steady-state security regions of power systems using optimization method," *Electr. Power Syst. Res.*, vol. 44, pp. 101–105, 1998.

[3] C. C. Liu, "A new method for the construction of maximal steady-state security regions of power systems," *IEEE Trans. Power Syst.*, vol. 1, no. 4, pp. 19–26, 1986.

[4] Z. Jizhong, *Optimization of power system operation*, vol. 53, no. 9. 2013.

[5] A. Capasso, A. Cervone, M. C. Falvo, R. Lamedica, G. M. Giannuzzi, and R. Zaottini, "Bulk indices for transmission grids flexibility assessment in electricity market: A real application," *Int. J. Electr. Power Energy Syst.*, vol. 56, pp. 332–339, 2014.

[6] J. Wu, U. Schlichtmann, and Y. Shi, "On the Measurement of Power Grid Robustness Under Load Uncertainties," in *2016 IEEE International Conference on Smart Grid Communications (SmartGridComm)*, 2016, pp. 218–223.

[7] A. A. Jahromi and S. Member, "On the Loadability Sets of Power Systems — Part I :," *IEEE Trans. Power Syst.*, vol. 32, no. 1, pp. 137–145, 2017.

[8] H. Yu, C. Y. Chung, and K. P. Wong, "Robust transmission network expansion planning method with Taguchi's orthogonal array testing," *IEEE Trans. Power Syst.*, vol. 26, no. 3, pp. 1573–1580, 2011.

[9] R. A. Jabr, "Robust transmission network expansion planning with uncertain renewable generation and loads," *IEEE Trans. Power Syst.*, vol. 28, no. 4, pp. 4558–4567, 2013.

[10] P. Wu, H. Cheng, and J. Xing, "The Interval Minimum Load Cutting Problem in the Process of Transmission Network Expansion Planning Considering Uncertainty in Demand," vol. 23, no. 3, pp. 1497–1506, 2008.

[11] J. Li, F. Liu, Z. Li, C. Shao, and X. Liu, "Grid-side flexibility of power systems in integrating large-scale renewable generations : A critical review on concepts , formulations and solution approaches," *Renew. Sustain. Energy Rev.*, vol. 93, no. March, pp. 272–284, 2018.

[12] F. B. Thiam and C. L. Demarco, "Transmission expansion via maximization of the volume of feasible bus injections," *Electr. Power Syst. Res.*, vol. 116, pp. 36–44, 2014.

[13] A. J. Wood and B. F. Wollenberg, *Power Generation, Operation, and Control*, 2nd ed. John Wiley and Sons Inc., 2012.

[14] H. Cetinay, F. A. Kuipers, and P. Van Mieghem, "A Topological Investigation of Power Flow," *IEEE Syst. J.*, vol. 12, no. 3, pp. 2524–2532, 2018.

[15] J. Lever, M. Krzywinski, and N. Altman, "Principal component analysis," *Nat. Publ. Gr.*, vol. 14, no. 7, pp. 641–642, 2017.

[16] "Index of data Illinois Institute of Technology." [Online]. Available: http://motor.ece.iit.edu/data/JEAS_IEEE118.doc. [Accessed: 20-Jun-2018].